# Scientific Objectivity and its Limits

## Richard Healey

## Abstract

Quantum theory is a fundamental part of contemporary science. But some recent arguments have been taken to show that if this theory is universally applicable then the outcome of a quantum measurement is not an objective fact. They motivate the more general reappraisal of the notions of fact and objectivity that I offer here. I argue that if quantum theory is universally applicable the facts about the physical world include a fact about each quantum measurement outcome. The physical facts may lack an ideal kind of objectivity but their more modest objectivity is all that science needs.

## 1. Introduction

Scientific procedures are designed to secure objectivity in part by minimizing the effects of individual differences in the ways scientists select, collect and report the data used to assess their theories. As the word suggests, at some level the data are supposed to be something anyone can recognize as simply objective facts—there for anyone to observe. Measurement outcomes provide data for a physical theory. Unless they are objective they support no objective scientific knowledge. So the outcome of a quantum measurement must be an objective physical fact. But recent arguments[1] have been taken to show that if quantum theory is universally applicable then the outcome of a quantum measurement is not an objective fact. They motivate the more general reappraisal of the notions of fact and objectivity that I offer here. If quantum theory is universally applicable the facts about the physical world include a fact about each quantum measurement outcome. They may lack an ideal kind of objectivity but their more modest objectivity is all that science needs.

      Section 2 explains the conclusion of one argument based on Leegwater's ([2018]) analysis of a *Gedankenexperiment*.[2] I sketch the argument itself in an Appendix. It shows how the assumption that quantum theory is universally applicable leads to the conclusion that not every quantum measurement in this scenario can have a unique objective outcome.

      Some may doubt the argument's conclusion because they believe quantum theory does not apply universally.[3] Others may doubt it because they believe quantum theory applies only

---

[1] Including those in (Bong *et al.* [2020]; Brukner [2017], [2018], [2020]; Cavalcanti [2020]; Frauchiger and Renner [2018]; Healey [2018]; Leegwater [2018]).

[2] I chose this because it may be stated compactly and requires fewer questionable assumptions.

[3] I take the universal applicability of quantum theory to rule out discontinuous physical change in a quantum state, either on measurement or spontaneously as in so-called 'collapse' theories (Ghirardi and Bassi [2020]).

in some preferred frame.[4] Everettians may readily accept the conclusion because they already assume multiple contrary outcomes, each observable in a different world. But they have something to learn about objectivity from the structure of 'worlds' in the scenario in which the argument is set and from the more general discussion in sections 3−6.

The argument impacts a collection of views on how to understand quantum theory that deny that a quantum state is what Bell ([2004]) called a *beable*—an element of physical reality, according to a theory in which it figures. Instead of describing or representing the physical condition or behaviour of a system, these views take a quantum state simply to yield Born probability measures over alternative possible events in which it may be involved (including measurements). That is roughly what Schrödinger ([1935]), QBists (Fuchs, Mermin and Schack [2014]), neo-Copenhagenists (Brukner [2017]) and pragmatists (Healey [2017a]), along with many other physicists (including Bohr?) take to be the constitutive function of the quantum state. Some readers may not realize how wide is the consensus on this view of quantum states since much recent philosophical discussion of quantum theory has focused on Everettian, Bohmian and physical 'collapse' interpretations/theories which take a quantum state to have a more than merely epistemic or prescriptive function.[5]

Consensus on this constitutive function soon dissolves into disagreement. Arguments mentioned in note 1 have prompted some (QBists) to emphasize the subjectivity of measurement outcomes, while others have concluded that they are relative rather than absolute. Here I focus instead on the consequences for objectivity and defend scientific objectivity against relativist and subjectivist threats.

Since experiments to date have placed no limits on the universal applicability of quantum theory Wheeler's 'radical conservatism' counsels us to assume there are none. In various *Gedankenexperimenten* extending that of Wigner's ([1961]) friend this implies that not all quantum measurements can have unique outcomes. This is a paradoxical conclusion in what I'll call the consensus view that the constitutive function of a quantum state is to yield probabilities for just such unique outcomes.

Section 3 considers attempts to resolve the paradox by appeal to the contextuality of sentences purporting to state facts about unique measurement outcomes. Appeals to contextuality of use fail to do so, but if the content of a sentence depends on the context at which it is assessed then the paradox may be avoided. Section 4 raises and answers the question as to how these sentences are assessable for truth or falsity only at an appropriate context. A sentence used in one context is assessable at another only insofar as quantum theory can be applied at both contexts at once in a wider context.

Section 5 applies and extends Huw Price's ([1988], [2003]) work on the notions of fact and truth to say why these notions may be limited in application by physical limits to communication while continuing to serve the functions that explain why we have them. I conclude by applying the lessons of the paradox to the world as it is—a world in which realizing the paradoxical scenario will forever remain beyond the powers of any agents. In the consensus view of our quantum world there are no transcendently objective facts about the outcomes of quantum measurements—nor indeed about the values of electric currents, nor even about the position of the apparatus in the laboratory. But these are immanently objective

---

[4] In Bohmian mechanics a universal wave-function evolves continuously in the preferred frame in which particle positions are taken to be distributed in accordance with the Born rule.

[5] For work by philosophers who take this consensus view seriously, see (Dascal [2020]; Dieks [2019a,b]; Evans [2020]; Healey [2017a,b]).

facts—facts expressed by sentences used to express truths when assessed at all actual contexts. Such facts provide all the objective data we need as scientists or those who rely on the objectivity of the knowledge they support.

## 2. Analyzing the Paradox

A set of plausible assumptions cannot all be taken to be true when applied in the scenario of the following *Gedankenexperiment*. A GHZ quantum state of three spin-½ particles is prepared. Alice, Bob and Charlie are in isolated labs when each measures the *z*-spin of a different particle at the same *t*-time in their common inertial frame. After each as has recorded their outcome at $t_1$, Eugene, Daniel, and Johnny (who are also in separately isolated labs stationary in that frame) each then measure a similar *X*-observable at the same *t*-time, recording their outcomes at $t_2$: Eugene measures $X_A$ on Alice's lab (including her particle, recording equipment and body), Johnny measures $X_B$ on Bob's lab, and Daniel measures $X_C$ on Charlie's lab; they record their outcomes at $t_2$. Alice's, Bob's and Charlie's measurements are pairwise spacelike separated; and so are Eugene's, Daniel's, and Johnny's. Though (absolutely) later than Alice's measurement, Eugene's measurement is also spacelike separated from Bob's and from Charlie's; Johnny's later measurement is spacelike separated from Alice's and from Charlie's; and Daniel's later measurement is spacelike separated from Alice's and from Bob's.

For this GHZ state[6], the Born rule yields joint probabilities for the outcomes of each triple of spacelike separated measurements. Let *a, b, c* number whether the *z*-spin of an atom was recorded as up (+1) or down (−1) by Alice, Bob, Charlie respectively; and let *u, v, w* number whether the relevant *X* observable on Alice's, Bob's, Charlie's lab was recorded as +1 or as −1 by Eugene, Johnny, Daniel respectively. The Born rule here implies that (with probability 1):

$$u.b.c = +1$$
$$a.v.c = +1 \qquad (1)$$
$$a.b.w = +1$$

But the Born rule also implies (with probability 1) that

$$u.v.w. = -1 \qquad (2)$$

Since the Born probability of any individual outcome other than ±1 is zero, each of *a, b, c* must (probability 1) be either +1 or −1. Multiplying equations (1) together gives $u.v.w. = +1$, in contradiction to (2).

This paradoxical conclusion follows from two broad assumptions: that there is a fact about the outcome of every (competent) quantum measurement and that quantum theory is universally applicable. But the content of each assumption requires clarification.

I'll take quantum theory to be universally applicable only if the evolution of an isolated system's quantum state is always unitary, with no physical 'collapses' or 'jumps',[7, 8] where this requirement holds in any special relativistic inertial frame when quantum theory is applied in Minkowski spacetime.

A system remains isolated while not interacting with any other system: any interaction would be marked by an interaction term in the Hamiltonian when quantum theory is applied to a supersystem containing that system. Isolation need involve neither spatial distancing nor sealing-off with barriers.

---

[6] If necessary after Lorentz-transforming this state to the relevant inertial frame.
[7] This is the Schrödinger picture state.
[8] That quantum theory is universally applicable in this respect is consistent with the results of experiments performed to date looking for violations of unitarity associated with so-called 'collapse' theories including those of (Ghirardi, Rimini and Weber [1986]; Penrose [2014]).

Unitary evolution of the quantum state is assumed to hold even for isolated laboratories including conscious experimenters and any system on which they perform a quantum measurement and record the outcome. And it is assumed to hold despite such an experimenter's reassignment of the quantum state of the system measured in her laboratory to reflect the measurement's outcome.

The Born rule is assumed to apply to each individual and joint measurement involved in the scenario of the *Gedankenexperiment*, and to yield the correct probabilities—here, as elsewhere.

With these clarifications, the assumption that quantum theory is universally applicable may be summarized as follows:

> Universality of Quantum Theory: Unitary quantum theory with the Born rule applies to every isolated system in any frame, including to the contents of an isolated laboratory in which a quantum measurement is performed.

The second assumption leading to paradox is that there is a fact about the outcome of every quantum measurement. Now every experimenter knows that some measurements fail to yield an outcome because something goes wrong: perhaps the detector is inefficient, the apparatus is disturbed, or the power goes out. This assumption needs to abstract from such real-life messiness without circular appeal to a quantum measurement's success. This may be done by modeling a quantum measurement as an interaction between a target system and an apparatus. I'll assume that all measurements in the paradoxical scenario are ideal Von Neumann measurements involving interactions that perfectly correlate initial eigenstates of the measured observable with final eigenstates of the corresponding apparatus system while leaving the initial system eigenstate unchanged. Consistent with the universal applicability of quantum theory, this permits a formulation of the assumption that there is a fact about the unique outcome of a quantum measurement (so modeled) with no appearance of circularity.

Applied to an initial superposed system state, an ideal Von Neumann measurement yields a final superposed quantum state not associated with the apparatus system's 'pointer' indicating any particular value of the measured magnitude. That is how the quantum measurement problem arises here. But no inconsistency arises on the consensus view that a quantum state does not represent the physical condition of systems to which it is assigned. The final (superposed) quantum state of target system and apparatus has no such representational role in the consensus view of section 1. Understood this way, a quantum state is associated not with values of magnitudes of the system, but with the possible values magnitudes of this or other systems may acquire as a result of a suitable interaction, importantly including measurement interactions.

The second assumption may now be stated as:

> Single Outcomes: Any quantum measurement has only one single outcome.

If quantum theory is universally applicable then it may be applied in the scenario of the *Gedankenexperiment* in which all six measurements there described are modeled as ideal von Neumann measurements. The second assumption implies that there is a fact about the outcome of the measurement by each of Alice, Bob, Charlie, Eugene, Johnny and Daniel.

The following sentences concern the outcome of Alice's measurement:

(A)   The outcome of Alice's *z*-spin measurement at $t_1$ is spin up.
(A)′  The outcome of Alice's *z*-spin measurement at $t_1$ is spin down.

(A) is true if the outcome of Alice's *z*-spin measurement at $t_1$ is spin up but false if the outcome of Alice's *z*-spin measurement at $t_1$ is spin down, in which case (A)′ is true. There is fact about the outcome of Alice's *z*-spin measurement if and only if exactly one of (A), (A)′ is true. I understand the occurrence of the word 'is' in (A) and (A)′ as tenseless: the sentence itself specifies the time of the outcome in Σ. The outcome may be ideally localized to a spacetime point assigned a time in any relevant frame. Just after that time Alice or one of her lab assistants may have used either (A) or (A)′ to state a fact about the unique outcome of the measurement. I follow Leegwater ([2018]) in assuming this is a fact only if the truth of that statement is not relative to an inertial frame. The outcome is recorded in the physical condition of a variety of objects in Alice's laboratory (where the pointer points, what is stored on her computer's hard drive, what is written in her notebook, and so on). While the description of a piece of recording equipment may be frame-relative, in each frame this condition serves to record the same outcome.

(A) and (A)′ have obvious analogs in sentences (B), (B)′; (C), (C)′ that Bob, Charlie respectively may use in his laboratory to state and characterize the fact that his *z*-spin measurement has an outcome. Eugene can use one of (U) or (U)′ to state and characterize the fact that his $X_A$-measurement had an outcome:

(U)   The outcome of Eugene's $X_A$-measurement at $t_2$ is +1.
(U)′  The outcome of Eugene's $X_A$-measurement at $t_2$ is −1.

(V), (V)′; (W), (W)′ are analogous sentences for Johnny, Daniel respectively.

The assumption that there is a fact about the outcome of each of these six quantum measurements then implies that exactly one of the 64 sentences of the form

(S)   A*& B*&C*&U*&V*&W*

may be used to truly state the outcomes of all six measurements. (An instance of schema (S) results from replacing A* either by (A) or by (A)′ and making analogous independent replacements of all the other starred letters). Moreover, this is a fact only if the truth of that statement is not relative to inertial frame. Here each conjunct of an instance of (S) is understood to be asserted in the laboratory where, and just after, the relevant outcome occurred.

A paradox arises here because applications of the Born rule imply that there is zero probability that any of the 64 sentences of the form (S) is true, and probability one that each is false. But there is a fact about the outcome of each of these six measurements only if exactly one of these 64 sentences is true. What kind of paradox is this?

A logical paradox occurs when a set of independently plausible assumptions implies a contradiction. We have arrived such a situation here only if the occurrence of a probability zero event in a finite event space implies a contradiction. The occurrence of such an event is often said to be impossible. But a contradiction arises only if we add the further assumption that a probability zero event does not occur, at least in this case. Is the occurrence of an event here consistent with its having probability zero?

This would be inconsistent if probability were a quasi-logical concept, as some (including Keynes ([1921])) have taken it to be. If probability is viewed as a kind of graded implication relation between premises and conclusion then probability one would correspond to logical implication, and probability zero to logical inconsistency. And if probability were a kind of graded disposition—a propensity—then a probability of one or zero would correspond to a sure-fire disposition, implying the occurrence or non-occurrence of an event. But neither of these views of probability seems adequate to the case of Born probabilities, and neither is adopted by most proponents of the consensus view of section 1.

Such views take probability assignments to have an epistemic or normative rather than a logical or metaphysical function. Some take a Born probability as a measure of an agent's actual coherent degree of belief, while others hold it up as a normative standard to which the degrees of belief of any agent who accepts quantum theory should conform. This distinction does not matter for present purposes, since they all take the paradox arising from the scenario of the *Gedankenexperiment* to be epistemic. Anyone who accepts the Born probabilities of that scenario should be certain that every sentence of the form (S) is false, while if she believes that there is a fact about the unique outcome of every quantum measurement then she should also be certain that exactly one such sentence is true.

### 3. Semantic Contextuality[9]

Suppose a sentence like (A) used to state the outcome of a quantum measurement were semantically contextual because what statement it expresses varies with its context of use. Then it may express a true statement when used in Alice's laboratory just after her measurement but a false statement when used in other contexts. Assume the same holds for each of the other eleven similar sentences that may be used to state the outcome of a quantum measurement in the scenario of the *Gedankenexperiment*. The paradox might then be resolved if, for each conjunct of a sentence of the form (S), there is a context in which the statement expressed by using a sentence of that form is true, while in no context can their conjunction be used to make a true statement.

(A) is not explicitly semantically contextual because it contains no indexicals such as 'I', 'now' or 'here'. Is it nevertheless use-contextual because its content depends implicitly on the context in which it is used? To resolve the paradox by appeal to such implicit use-contextuality one would have to parametrize contexts of use so that (A) may express a truth when used in Alice's lab just after her measurement but a false statement when used in some other relevant context. Contexts of use have been parametrized by Kaplan ([1989]) and others as ($t$, $s$, $p$, $w$), where $t$ indicates the time and $s$ the place of use by person $p$ in world $w$. Variation in none of these parameters generates a context of use in which (A) makes a false statement, if (A) expresses a true statement when used in Alice's lab just after her measurement. Variation in $w$ is irrelevant because the paradox is set in a single world $w$, assumed to be possible consistent with the universal applicability of quantum theory. Alice's lab assistant states the same truth as Alice when using (A) in the same circumstances, so variation in $p$ would not yield a false statement. Like all the other 11 sentences describing the outcome of a quantum measurement in this scenario, (A) itself specifies both the laboratory in which that outcome occurs and the time when it occurs. There is no reason to suppose that any of these sentences expresses a statement with a different truth-value when used at different times or places. In the absence of any further relevant parameters marking a context of use, appeal to the use-contextuality of sentences like (A) fails to resolve the paradox.

---

[9] Semantic contextuality arises whenever the truth-value of a sentence is sensitive to the circumstances in which it may be used or assessed. The term 'contextual' has a different sense in quantum foundations to refer to a theory in which the outcome of measuring an observable may depend on what other compatible observables are measured along with it. This section goes some to way toward connecting these apparently unrelated senses.

To resolve the paradox we need to appeal to a further kind of context which (following MacFarlane ([2014])) I call a context of assessment. MacFarlane's motive for introducing contexts of assessment is very different from mine. He uses them as part of a semantic proposal for a natural language like English that would make room for truth-relativism—the possibility that a sentence used to make a statement with fixed content may be correctly assessed as true at one context but false at another. No friend of scientific truth-relativism, I'll use contexts of assessment first to resolve the paradox and then to explore the consequent limits of scientific objectivity. The key idea will be that a statement assessed as true at one context cannot be assessed either as true or as false at some other context.

I'll take a context of assessment to involve a spacetime region such as the regions in which measurements occur in the paradoxical scenario. I call this context of assessment the 'environment' of such measurements. It will take some time to say just what kind of environment this is. But notice already that this context of assessment requires no specification of any person $p'$ (such as Alice) or world $w'$ other than that in which measurements are made, actually or in a *Gedankenexperiment*.

At least in the paradoxical scenario, every application of quantum theory is to a system or systems in a particular environment. As is now recognized by a range of views on how the theory is best understood, any legitimate application of quantum theory's probabilistic algorithm requires that a system's quantum state should be robustly decohered through a process in which the system interacts with other systems in its environment. I'll say that a 'quantum event' occurs when a system's magnitude takes on a value in its environment. A necessary condition for the occurrence of a quantum event involving a system is that there be a process in the system's environment that can be modeled by the robust decoherence of states in a system's 'pointer basis', each associated with a different value of that magnitude. This is not a sufficient condition. In the consensus view, quantum theory itself cannot explain the event in which the magnitude takes on one rather than another value. But quantum events are observed to occur, and application of the Born rule correctly predicts the probability for a magnitude to take on one value rather than another when one does.

An interaction suitable to function as a quantum measurement may (and often does) have an outcome in an environment outside any laboratory with no agent present. All that is required is for a quantum model of the interaction to imply robust decoherence of system states in some pointer basis. In a measurement, a quantum event occurs as a corresponding magnitude takes on a value. The outcome $o$ of a quantum measurement is typically marked by the occurrence of many separate such quantum events in different systems, but we can count their fusion also as the single composite quantum event $o$ that occurs in the spacetime region they collectively occupy.

An '$M$-decoherence environment $E$' of a quantum event $e$ in which magnitude $M$ takes on a value in a system is a region of spacetime $R_E$ that includes the region where $e$ occurs, together with physical processes in $R_E$ that can be modeled by robust decoherence of that system's states in a 'pointer basis' associated with different values of $M$. Such decoherence is robust in the sense that once a process starts the system's reduced state remains very nearly diagonal in the pointer basis throughout $R_E$. I'll call these the '$M$-decoherence processes' of the system in environment $E$. A 'decoherence environment' is an $M$-decoherence environment for some magnitude and system. A decoherence environment $F$ 'compatibly extends' a decoherence environment $E$ if and only if (for each $M$) all the $M$-decoherence processes of systems in $E$ are also $M$-decoherence processes of systems in $F$ (so $E$ compatibly extends itself). Two decoherence environments are 'distinct' just in case no process occurs in both their regions, and 'mutually isolated' if and only if they are distinct and there is no interaction between any process in $E$ and any process in $F$. A decoherence environment is 'isolated' if and only if it is mutually isolated from all distinct decoherence environments. Note that decoherence environments $E$, $F$ may be mutually isolated if $R_E \cap R_F \neq \emptyset$, or even if $R_E = R_F$.

A 'context of assessment' for each of the twelve sentences purporting to state the unique outcome $o$ of a quantum measurement in the paradoxical scenario of section 2 is a decoherence environment $E$. $E$ is a decoherence environment for which $o$ occurs in $R_E$ if and only if $E$ is an $M$-decoherence environment for every magnitude $M$ that takes on a value in a quantum event that is part of $o$. The interior of Alice's laboratory throughout her $z$-spin measurement occupies the region $R_A$ of the 'primary' context of assessment $E_A$ for a sentence (A): (A) is assessed as true at $E_A$ if Alice's outcome is spin up, but false if it is spin down. Each of the other eleven sentences that figure in the paradoxical scenario may be assessed at its own primary context of assessment.

Since $R_A$, $R_B$ are spacelike separated there can be no interaction between a process in $R_A$ and a process in $R_B$: so $E_A$, $E_B$ are mutually isolated. But there is a wider context of assessment $E_{AB}$ in region $R_A \cup R_B$ that is a compatible extension both of $E_A$ and of $E_B$. Since $z$-spin measurements in $R_A$, $R_B$ are compatible, the Born rule may be applied to joint $z$-spin measurements in $R_A \cup R_B$, so each of (A), (B) may be assessed as true or as false at $E_{AB}$. Analogously, $R_C$ is spacelike separated from $R_A$, $R_B$, so there is a context of assessment $E_{ABC}$ in $R_A \cup R_B \cup R_C$ that compatibly extends $E_{AB}$ and $E_C$ at which (A), (B), (C) may each be assessed for truth-value.

Eugene's measurement also occurs at spacelike separation from those of Bob and Charlie, and there is a context of assessment $E_{UBC}$ for (U), (B), (C) that compatibly extends all of $E_U$, $E_B$, $E_C$. But there is no context of assessment at which each of (A), (B), (C), (U) may be assessed for truth-value, because $E_A$, $E_B$, $E_C$, $E_U$ have no common compatible extension. This is because no decoherence environment $E_{AU}$ compatibly extends both $M$-decoherence environment $E_A$ for a magnitude $M$ that takes on a value in region of spacetime $R_A$ to record Alice's outcome and $N$-decoherence environment $E_U$ for a magnitude $N$ that takes on a value in a region of spacetime $R_U$ to record Eugene's outcome. Decoherence in a region of spacetime $R_A \cup R_U$ including both Alice's and Eugene's outcomes may privilege an $M$ basis of 'pointer states' or it may privilege an $N$ basis. But because the observable that Alice measures is incompatible with the observable that Eugene measures there is no common basis of pointer states that can serve to record both their outcomes at once. Essentially, Eugene's measurement removes all traces of Alice's measurement outcome; Johnny's measurement similarly erases Bob's, and Daniel's erases Charlie's.

In the paradoxical scenario there are five contexts of assessment $E_{ABC}$, $E_{ABW}$, $E_{AVC}$, $E_{UBC}$, $E_{UVW}$, each of which compatibly extends the primary context of assessment for three of the measurements in that scenario. But there is no compatible extension of more than three of these primary contexts. It follows that no sentence of the form (S) can be assessed as true or as false at any context of assessment that compatibly extends the primary context of assessment for each of its conjuncts.

An experimenter in each laboratory does assess the corresponding conjunct for truth or falsity by recording the outcome of the measurement in that laboratory. For example, there is a fact about the outcome of Alice's $z$-spin measurement if and only if (A) or (A)′ is true but not both: and Alice assesses the truth-value of (A) at context $E_A$. Leegwater ([unpublished]) required a unique outcome of a quantum measurement to be independent of perspective. This can now be understood to require that Alice's measurement has a unique outcome only if (A) has the same truth-value not only in the perspective provided by context $E_A$ but also in the perspectives provided by its compatible extensions $E_{ABC}$, $E_{ABW}$, $E_{AVC}$: that (B) has the same truth-value at all analogous compatible extensions of context $E_B$ while (C) has the same truth-value at all analogous compatible extensions of $E_C$: that (U) has the same truth-value at $E_U$ as at $E_{UBC}$ and at $E_{UVW}$, while (V), (W) each has the same truth-value at its primary context of assessment as at its analogous compatible extensions. But the paradox shows that it is irrational to believe these conditions can all be satisfied at once.

## 4. Limits on Assessment

If unitary quantum theory is universally applicable, then there is no perspective-independent fact about the outcomes of the six measurements performed in the paradoxical scenario. This is true even if compatible extensions of each outcome's primary context of assessment offer the only available perspectives. But there are contexts of assessment such as $E_U$ for a sentence like (A) that do not compatibly extend its primary context of assessment. Should (A) not receive the same truth-value at $E_U$ as at $E_A$? Why should contexts of assessment be restricted to decoherence environments: does a region of spacetime in which no decoherence processes occur not also offer a perspective on the outcome of Alice's measurement? We need to understand how there can be limits to the assessment of sentences such as (A) and how this places limits on scientific objectivity.

Thomas Nagel's ([1986]) famous metaphor of the view from nowhere suggests an ideal of objectivity according to which a thought or statement expresses an objective truth if and only if its truth depends only on how things are in the world and has nothing to do with any contexts at which it might be assessed as true or as false. Any statement that meets this ideal may be called 'transcendently objective'. The paradox shows that a sentence like (A) about the outcome of a quantum measurement cannot be used to make a transcendently objective statement about the world—to state what Brukner ([2018]) called a fact of the world. It shows that there can be no transcendently objective facts about the outcomes of these quantum measurements. It shows that there are not even perspective-invariant facts about these outcomes, where a fact is perspective-invariant if and only if it may be expressed by a statement that is true when assessed from every perspective.

Facts about the outcomes of quantum measurements in the paradoxical scenario are perspectival—what they are depends on the context of assessment because a statement assessed as true at one context cannot be assessed either as true or as false at others. I caution against calling these 'relative facts', because to do so falsely suggests that a statement about the outcome of a quantum measurement may be assessed as true at one context but as false at another. Quantum theory involves no such truth-relativism. It is factuality, not truth-value, which is perspectival in the paradoxical scenario. The problem is to understand how this is possible. This is a problem because commonly held preconceptions apparently rule out any notion of a perspectival fact. Reflection on another case that presents limits to assessment may help to loosen their hold on the imagination.

Consider the sentence (H): It is five o'clock here. (H) is use-contextual: 'here' marks a spatial index and the present tense marks an implicit temporal index. Once these are fixed by specifying the time and place of use, (H) may seem assessable as true or false (in the actual world) independent of context. But, as Wittgenstein ([1953], §350) recognized, a use of (H) on the sun has no readily assessable meaning or truth-condition on earth. The sun is in no terrestrial time-zone, terrestrial clocks cannot exist on the sun, and assessment of (H) at a terrestrial time and place presupposes a non-relativistic notion of absolute simultaneity.

Whether (H) is assessable at a terrestrial time and location is contingent on the availability of a physics capable of jointly modeling the sun and earth within a unified (spatiotemporal) framework. Both Newtonian and relativistic physical theories provide such a framework, though relativity leaves some latitude in how to model distant simultaneity. Our physics provides a theoretical framework that may be used to render (H) assessable here and now on earth.

But what physics provides it can also take away. Although we have reason to believe it does not represent the structure of our actual universe, Gödel's ([1949]) model of general relativity illustrates this possibility. In this model there are closed timelike curves: indeed, any two points of the spacetime lie on such curve, so each event occurs both before and after itself in the curve's time ordering. With no cosmic time in the Gödel universe it is impossible to introduce a corresponding notion of distant simultaneity, however arbitrary. The truth-value of (H) as used at some point in the Gödel universe cannot be assessed at any other point. The sentence (H) could not be used to state a transcendently objective or even perspective-invariant fact in the Gödel universe.

Quantum theory now has an unbroken record of successful applications for close to a century. It has been successfully applied in the realm of the very large and the very small, over a huge range of energies, and to both simple and increasingly complex systems. It is appropriate to let our intuitions be guided by these successful applications when considering the limits on the assessment of sentences that may be used to state their results, including the outcomes of quantum measurements.

Applications of the Born rule are restricted to so-called compatible properties of a system—properties that quantum theory associates with non-commuting projection operators. A variety of no-go results rule out any extension to an assignment of joint probabilities to incompatible properties. If we are to be guided by our best physical theories, we should conclude that sentences ascribing incompatible properties to a system cannot be assessed together as true or as false. This is a significant restriction on contexts of assessment. It is met by sentences used to assert the outcome of a quantum measurement jointly observing a set of properties, since such properties are observable together only if they are compatible.

Contexts of assessment need not be confined to occasions on which these properties are actually observed in a quantum measurement: outcomes of quantum measurements are not observer-dependent. Sentences about properties of magnitudes may also be jointly assessed even though no experimenter performs a quantum measurement.[10] They may be jointly assessed at any *M*-decoherence environment in which distinct *M*-values uniquely correspond to possession of contrary properties. But since no application of quantum theory requires or permits joint assessment outside of such an *M*-decoherence environment we should be guided by the theory in refusing to countenance any other contexts of assessment for sentences ascribing properties to quantum systems.

## 5. Facts and the Limits of Truth

Even if section 4 showed how there can be limits to the assessment of statements for truth and falsity, it did not say why there should be such limits. To understand why we need to look more closely at the notion of truth.

---

[10] So we should freely accept the obligation mentioned by Bell ([2004], p. 216) 'If the theory is to apply to anything but highly idealized laboratory operations, are we not obliged to admit that more or less "measurement-like" processes are going on more or less all the time, more or less everywhere?' But his follow-up question 'Do we not have jumping then all the time?' should be answered 'No', assuming these processes may be modeled by quantum decoherence with no jumping quantum states, in accord with the universal applicability of quantum theory.

Facts and truth are two sides of a coin. When Alice utters (A) after measuring her particle's $z$-spin she thereby takes herself to make a true statement. It is hard to think of clearer examples of statements of fact than scientists' reports of the results of their own observations (which does not mean every such sincere report is true—science builds on fallible foundations). To break out of the tight circle connecting truth and fact it is tempting to seek an analysis, of one notion or the other. But for a pragmatist there is a prior question: What is the point of these notions? Why do we have them? What good are they, and why can't we do without them?

In his book Price ([1988]) argues against the analytic approach to the concept of a statement of fact in the first part while offering an explanation of why we have the notion of truth in the second. The explanation helps one to understand not only why the notion of truth is important in many domains of thought but also why there are limits to its applicability. It thereby helps one to appreciate the limits of the notion of a fact in general (and of an objective fact in particular). Section 4 gave conditions under which certain statements could be assessed for truth or falsity. Given the function of the notion of truth, there is no point in applying the notion unless those conditions are met. This explains why these statements should be assessed for truth or falsity only under those conditions.

In a deflationary view, the notion of truth has only a kind of book-keeping function. Rather than repeating a formula like 'Tarski said/believed that snow is white, and snow is white' for each of Tarski's indefinitely many statements/beliefs, the truth predicate permits the simple generalization 'Everything Tarski said/believed is true'.[11] But while endorsing the deflationary idea that truth is not a substantial property, Price ([1988]) takes the notion of truth also to play a more socially significant normative role. By acknowledging truth as a norm of discourse, members of a speech community are motivated to engage in the kind of reasoned argument among members whose mental states initially differ that tends to align their mental states to the benefit of all. Such benefits accrue for mental states with what Price calls the Same Boat Property (SBP):

> A class of mental states have the SBP if their typical behavioural consequences are such that their behavioural appropriateness, or utility, is predominantly similar across a speech community. If a mental state has the SBP, then if it is appropriate for any one of us, it is appropriate for all—we are all in the same boat. (Price [1988], p. 152)

To probe the limits of a useful notion of truth in the paradoxical scenario we need to generalize the SBP so that it concerns the intentional states of a wider collection of agents than those composing a single human speech community. Here is a formulation of a Generalized Same Boat Property (GSBP)[12]:

---

[11] Since it follows from this deflationary view that (A) is true if and only if the outcome of Alice's $z$-spin measurement at $t_1$ is spin up, endorsing that equivalence does not commit one to any more inflated correspondence theory of truth: nor does acknowledging that the notion of truth has an additional social normative function.

[12] The GSBP offers neither a notion of truth nor an explanation of why truth is a useful notion. It simply characterizes a class of intentional states in Dennett's ([1989]) sense. These include many human mental states but perhaps also intentional states of cognitively sophisticated non-human social agents, conscious or otherwise. Public expression of a belief-like intentional state in this class amounts to a statement. Subjecting such statements to the norm of truth will tend to benefit at least the group.

> A class of intentional states have the GSBP if their typical behavioural consequences are such that their behavioural appropriateness, or utility, is predominantly similar across a collection of social agents. If an intentional state has the GSBP, then if it is appropriate for any agent in the collection, it is appropriate for all—they are all in the same boat.

A collection of social agents may be partitioned into mutually isolated groups where by expressing its own intentional states a member of one group is able to affect those of other members of its group, but unable to affect those of members of other groups. As long as these groups remain mutually isolated, expression of intentional states by members of one group cannot effectively align the intentional states of all agents in the collection. The norm of truth can fulfill its social function within each group, but not across the lines separating groups.

There may be fundamental physical barriers to communication. Even if there are societies of intentional agents over our cosmic event horizon we will never be able to affect each other's intentional states. There can be no communication between scientists in different isolated laboratories in the scenario of section 2. In each case we have a collection of agents segmented into physically and therefore socially isolated communities. So we might expect difficulties in applying a universal notion of truth to their statements, and corresponding limitations on the notion of a fact. But there is an important difference between the cases.

Current general relativistic models of the large scale structure of the universe give sense to assessment here and now of sentences about when an event occurred over our cosmic horizon, by incorporating a global cosmic time that crosses the horizon and thereby establishes a theoretical connection between events on opposite sides. Locating these events within a single theoretical model makes it significant and sometimes possible to assign a truth-value to other sentences also. Our best theories tell us that the cosmic microwave background radiation over our cosmic event horizon closely approximates the same spectrum of black body radiation, that space-time still has a metric satisfying Einstein's field equations, that there are galaxies of stars, and so on. Our justified confidence in our best theories dictates belief that there are non-perspectival facts about events over our cosmic event horizon just as it dictates belief in the existence of that horizon.

Contrast this with the paradoxical scenario of section 2. Here our best theory (quantum theory) connects contexts of assessment only if the theory may be applied in a compatible extension of all those contexts. For example, (A) may be assessed at $E_B$ and (B) at $E_A$ only because $E_{AB}$ compatibly extends each of $E_A$, $E_B$. After Eugene's measurement there is no context at which any agent can assess both (A) and (B). Alice and Bob can never share the outcomes of their experiments. Immediately after performing their experiments these are at spacelike separation: later they are effectively erased by Eugene's and Johnny's measurements. But this does not prevent the application of quantum theory at $E_{AB}$ that requires joint assessment of (A), (B) at that context even though no single experimenter can verify both outcomes. Alice's verification of the truth-value of (A) at $E_A$ thereby verifies it also at $E_{AB}$: but neither Alice, Bob nor anyone else can verify a truth-value assignment to both (A) and (B) at $E_{AB}$, even though the applicability of quantum theory requires their joint assessability at that context.

In a standard experiment verifying or exploiting violations of Bell inequalities, the outcomes of measurements at different locations can be shared later because their decoherence environments do not remain isolated but fuse into a single subsequent decoherence environment. In that fused environment many experimenters can verify the outcomes of all these measurements. We have confidence in the application of quantum theory to spatially separated measurements on entangled systems only because very many such verifications on a wide variety of systems assigned entangled quantum states have yielded statistics confirming joint probabilities predicted by the Born rule.

The features of the paradoxical scenario that distinguish it from a standard experiment manifesting non-classical patterns of correlation would also make it extraordinarily difficult to realize in an actual experiment. Preparing the entangled GHZ state would be difficult, but not beyond present technical capabilities. While each individual measurement by Alice, Bob or Charlie would not be hard to realize, it would be a challenge to ensure that measurements of any pair occur at spacelike separation. But there are two reasons why the paradoxical *Gedankenexperiment* is so far beyond the bounds of practicality that neither we nor any other community of agents will ever be able to perform it.

The first reason is that it is and seems likely always to remain far beyond our technical capabilities to perform any of the individual measurements by Eugene, Johnny and Daniel. It is easy formally to write down a unitary transformation that would model the interaction Eugene (for example) would have to apply to measure $X_A$ on Alice's laboratory together with the spin ½ atom she has measured. But the operations needed to implement this interaction would need to be applied not just to the large scale features of Alice's lab but precisely and in a very specific way at the level of all its microscopic constituents, including the atoms and ions composing the bodies of its occupants. It is hard to imagine how any agent could ever perform these operations. If performed, they would effectively remove any record of Alice's outcome: if Eugene were to observe her lab immediately after he would be equally likely to find multiple concordant apparent records of a spin up outcome or of a spin down outcome, whatever Alice had actually found. A recent paper (Aaronson, Atia and Susskind, [unpublished]) shows that this is a generic feature of measurements sensitive to such macroscopic superpositions.

The more fundamental reason why the scenario of section 2 will remain only a thought-experiment has to do with the assumption of physical isolation. The scenario assumed that a system could be isolated from the decohering effect of its external environment, in which case its environment corresponds to its own internal decoherence. But decoherence spreads extraordinarily widely extremely fast. Even in interstellar space a lab could not be isolated from the decohering influence of its external environment, including starlight, the all-pervasive 2.7 degree Kelvin microwave background radiation and even gravitational waves.

This second reason has implications for the objectivity of measurement outcomes. The practical impossibility of confining decoherence within any lab implies that experimenters are in practice always in the same boat: their intentional states concerning outcomes of their respective quantum measurements have the GSBP no matter how hard they may try to isolate their separate labs. Insofar as it is only physically localized agents that can perform quantum measurements, it is not just we humans that are inevitably always in the same boat: so would be any aliens or androids that may perform quantum measurements and acquire intentional states noting their outcomes. We all share a single decoherence environment and so we can reasonably expect to be able to reach agreement on the outcomes of any quantum measurements that can ever actually be performed. A notion of truth suitable for any actual community is universal enough, and yields all the objectivity we need as scientists and can hope for as people.

## 6. Scientific Objectivity Reassessed and Reasserted

The scenario of section 2 exhibits limits to truth and factuality, and thereby threatens the objectivity of science: there can be no transcendently objective facts about the outcomes of those quantum measurements. But since realizing that scenario is so far beyond our abilities it is not obvious what this implies about the kind of objectivity that matters for science. To address this question consider why its realization is so far beyond our abilities.

No system or laboratory can be completely isolated from the effects of environmental interactions associated with decoherence of its quantum state. It is decoherence of the right sort that makes quantum measurements possible. But even very weak and ineliminable interactions between a laboratory and its external environment would combine with the interaction needed to measure an $X$-magnitude involving that laboratory so as to effectively decohere its state in a model and thereby destroy a delicate correlation like $u.v.w. = -1$ that led to the paradox. To put it metaphorically, the paradoxical scenario will forever remain merely a thought-experiment because realizing it would require an impossible combination of the absence of decoherence external to each laboratory with the presence of just the right kind of decoherence within that laboratory. This is not just an impossible combination for human scientists to create: it is beyond the powers of any system complex enough to constitute an agent capable of noting the outcome of a quantum measurement as a subsystem of a laboratory system. So no alien, android, or quantum computer running an advanced AI program could play the role of an observer in realizing the paradoxical scenario.

In the view from nowhere, the measurements of Alice, Bob, Charlie, Eugene, Johnny and Daniel could not all have had unique outcomes. But transcendent objectivity is a metaphysical ideal, not a presupposition of successful science. Although its content may be diminished, a claim about the outcome of a single or joint measurement by the agents in the *Gedankenexperiment* is still objective in several senses. First, statements about measurement outcomes are not subjective: they are not about, and not relative to, the epistemic state or consciousness of the one who makes them. Secondly, any agent who has accepted quantum theory can adopt the perspective of any of Alice, Bob or their friends in the thought-experiment and understand, use and (hypothetically) assess the truth-value of statements about the outcome of a measurement from that perspective. Third, and most importantly, there are no completely isolated labs, and even if there were there is no conceivable practical way of performing the interactions needed to implement an $X$-measurement on such a lab and its contents. In all practically realizable situations scientists and all other agents now, and always will, share a single perspective because they inhabit a single environmental context.

This third sense is worth a name of its own. I'll call a statement 'immanently objective' if it expresses a truth at all actual contexts of assessment. The measurement outcomes in the paradoxical scenario would not be even immanently objective: Alice's statement of her outcome would be evaluated as true at her environment but not at Eugene's, for example. But the thought-experiment cannot be performed because quantum decoherence is pervasive in the only environment actually shared by all agents. This provides the only actual context of assessment for statements about the outcomes of quantum measurements. So these are all immanently objective—objective enough to provide data that supports quantum theory.

So far I have focused on the objectivity of measurement outcomes. But what I have said applies much more widely. Most if not all statements about the physical world may be recast as statements about the values of magnitudes. These include statements about settings of switches and knobs on experimental equipment and currents in coils, but also statements about the behavior of a mouse, the height of a plant, the structure of an FMRI image, and even where things are in and outside the lab. One can adopt the consensus view of quantum theory while taking such statements to be just as objective as science needs them to be, even though no statement about the value of a magnitude is transcendently objective: and since we share a single decoherence environment with all agents it is not just human science whose objectivity is not threatened by the paradox. The conceptual distinction between transcendent and immanent objectivity has no conceivable practical scientific significance, though it is of great philosophical interest.

# Appendix

Here I sketch the argument whose conclusion is explained in section 2. The scenario unfolds in a *Gedankenexperiment* involving a triple of spin ½ atoms initially assigned an entangled GHZ state. Measurements are made not only on these systems but also on entire isolated laboratories in which those measurements are made on them. Since Wigner ([1961]) is credited with first publicizing the possibility of performing a quantum measurement on an entire isolated laboratory containing his friend[13], this may be called a scenario where Greenberger, Horne and Zeilinger meet Wigner's friend—the title of a preprint by Leegwater ([unpublished]) whom I credit with this argument.

A.1. The scenario of Wigner's friend

Wigner ([1961]) considered a *Gedankenexperiment* in which his friend (whom I'll call Alice) is confined to a physically isolated laboratory while performing a quantum measurement on a system (I'll take this to be of the *z*-spin of a spin ½ atom initially assigned a spin-up quantum state of *x*-spin). Applying the Born rule of quantum theory to this state, Alice should expect a unique outcome of either *z*-spin up or *z*-spin down, each with probability ½. Eugene initially remains outside the laboratory without interacting with it. Assuming that quantum theory is applicable to any isolated system no matter how large or complex[14], Wigner (and his namesake Eugene) assigns a quantum state to the entire contents of Alice's laboratory (including her body as well as the atom and all her measuring and recording apparatus). He further assumes that this state evolves unitarily during Alice's measurement since her laboratory then remains isolated. After Alice's measurement this state will therefore be an entangled superposition, one component of which is often thought to correspond to Alice's outcome up (as observed by Alice and multiply recorded in her laboratory) while the other corresponds to Alice's outcome down (as similarly observed by Alice).

    The tension between the assumption that unitary quantum theory is universally applicable and the assumption that every quantum measurement has a unique outcome is already apparent. But it may be relieved in this case by blocking the inference from the superposed state to absence of a unique outcome. Despite assigning a superposed state to her laboratory, Eugene can consistently maintain that Alice's measurement had a unique outcome before breaking the isolation by opening the door to have a look for himself. Insisting on a unique outcome even with such a superposed state is not an *ad hoc* move on what I called the consensus view of the function of a quantum state. On this view the scenario of Wigner's friend yields no paradox.

---

[13] A referee kindly noted that Everett discussed essentially the same scenario in his later published 1956 thesis.
[14] This follows from the assumption that quantum theory is universally applicable, as further analyzed in section 2.

A.2. The scenario of Greenberger, Horne and Zeilinger

One consequence of Bell's theorem ([1964], [2004]) is that the probabilistic predictions of any local hidden variable theory are inconsistent with those of quantum theory in certain circumstances. The quantum predictions have now been thoroughly verified in a wide range of these circumstances. Greenberger, Horne and Zeilinger ([1990]) extended Bell's theorem to a situation involving only extremal (0 or 1) probabilistic predictions. They considered an entangled state of three quantum systems now known as a GHZ state. To explain their argument I'll consider a scenario involving a GHZ state of three spin ½ atoms (see equation (13) of Leegwater ([unpublished])).

The Born rule predicts that this state has a number of interesting properties:
(1) If the *y*-spin of each atom is measured simultaneously the outcome will (with probability 1) be either all spin up or all spin down, and each outcome is equally likely (probability ½).
(2) If the *x*-spin of each atom is measured simultaneously there are four possible (non-zero probability) outcomes; but in each case if we assign +1 to a spin up outcome and −1 to a spin down outcome then the product of these three numbers will (with probability one) be −1.
(3) If the *x*-spin of one atom is measured simultaneously with measurements of the *z*-spins of the other two there are four possible (non-zero probability) outcomes; and if up/down outcomes are assigned +1/−1 as in (2), then the product of these three numbers will (with probability one) be +1.

Suppose that a measurement of a component of spin merely reveals whether that component had spin up or down when it was measured. Then we can assign numbers to these properties just as we assigned numbers to the outcomes of a corresponding spin measurement. Let *a, b, c* number whether the *x*-spin of the first, second, third atom was up or down; and let *u, v, w* number whether the *z*-spin of the corresponding atom was up or down. Then (2) implies that $a.b.c = -1$, while (3) implies that

$$a.v.w = +1$$
$$u.b.w = +1$$
$$u.v.c = +1$$

Multiplying these last three equations together it follows that $a.b.c = +1$, in contradiction to what (2) implied. So, with probability one, the Born probabilities predicted by the GHZ state are incompatible with the assumption that each measurement merely revealed the value of the measured spin-component. Experiments have verified the Born probabilities for similar GHZ states (Pan *et al.* [2000]).

A.3. Wigner's friend meets GHZ

Though striking, neither of the scenarios described so far is paradoxical. Each may be rendered compatible with the assumptions that unitary quantum theory is universally applicable and that every quantum measurement has a unique outcome. We arrive at a paradox only by suitably combining these scenarios into a third, more complex scenario.

The idea is to substitute the assumed outcomes of actual quantum measurements for the hidden states assumed in the GHZ scenario. These will be quantum measurements on three spin ½ atoms assigned a GHZ state, and they will be performed in three mutually isolated laboratories. Each laboratory will be assumed also to be isolated from everything else except during the performance of a measurement on that laboratory. This last measurement will be analogous to a measurement Eugene might have performed to verify the quantum state he has assigned to his friend's lab and its contents in the scenario of A.1.

So consider three experimenters Alice, Bob and Charlie, each confined to a physically isolated laboratory to perform one measurement of spin component on a single spin ½ atom. Assume that these three atoms are assigned the GHZ state for which the Born rule correctly predicts probabilities satisfying conditions (1)–(3) specified in A.2. Note that those conditions concerned simultaneous measurements. In the combined scenario it will be important to be explicit about what that means, since the scenario will be set in a relativistic spacetime that is assumed to have the structure of Minkowski spacetime, at least in the region where it unfolds. The laboratories of Alice, Bob and Charlie are all stationary in inertial frame $\Sigma$ in which they perform their measurements simultaneously.

The universal applicability of quantum theory here implies that the Born rule may be correctly applied to a set of (mutually compatible) measurements that are simultaneous in any inertial frame. Eugene, Johnny and Daniel remain stationary in frame $\Sigma$. Eugene measures $X_A$ on Alice's laboratory, Johnny measures $X_B$ on Bob's laboratory, and Daniel measures $X_C$ on Charlie's laboratory. These three measurements also occur simultaneously in $\Sigma$.

Following Leegwater ([unpublished])[15], the assumption that a quantum measurement has a unique outcome may now be stated as:

Single Outcomes: Any quantum measurement has only one single outcome.

This assumption applies to each of the six specified measurements. But it will turn out that after all of them have been completed the only surviving records are of the outcomes of the measurements by Eugene, Johnny and Daniel.

Between $t_0$ and $t_1$ in $\Sigma$ each of Alice, Bob and Charlie measures the $z$-component of spin of the atom in his or her laboratory and records its unique outcome inside the laboratory without breaking isolation by communicating this outcome outside. Eugene, Johnny and Daniel make their measurements between $t_1$ and $t_2$ in $\Sigma$. Eugene measures $X_A$ on Alice's laboratory together with the spin ½ atom she has measured. Johnny (Daniel) measures $X_B$ ($X_C$) on Bob's (Charlie's) laboratory. An $X$-magnitude is the analog in the argument based on this scenario of $x$-spin in the GHZ scenario. But, unlike the $x$-spin of an atom, it would be extraordinarily difficult to measure an $X$-magnitude: it may be thought to encode detailed correlations between a spin ½ atom and the atomic structure of the laboratory that contains it.

Assuming their labs are far enough apart in $\Sigma$ there will be an inertial frame $\Sigma'$ in which the measurements by Bob and Charlie between $t_1{'}$ and $t_2{'}$ are simultaneous with Eugene's but after Alice's. The quantum state assigned at a $t'$ time just prior to these simultaneous measurements will be an analog[16] of the state in the original GHZ scenario that implied an instance of condition (3) of A.2. We can number the possible outcomes of Eugene's, Bob's and Charlie's simultaneous measurements, each with +1 or −1 as before. Let Bob, Charlie's and Eugene's outcomes be numbered $b, c, u$ respectively. Then the analog to condition (3) implies in this case that $u.b.c = +1$.

There is another inertial frame $\Sigma''$ in which it is not Bob's but Johnny's measurement that is simultaneous with those of Alice and Charlie. Numbering Johnny's outcome $v$ the analog to condition (3) implies that $a.v.c = +1$. Finally it is Daniel's not Charlie's measurement that is simultaneous with Alice's and Bob's in $\Sigma'''$, from which we infer that $a.b.w = +1$, where $w$ numbers Daniel's outcome. Putting these equations together we have

$$u.b.c = +1$$
$$a.v.c = +1$$
$$a.b.w = +1$$

---

[15] Leegwater ([unpublished]) also requires the outcome to be independent of 'perspective', but he does not say what that is or how it differs from a frame. This will prove significant in section 3.

[16] As Leegwater ([unpublished]) notes, the relativistic transformation from $\Sigma$ to $\Sigma'$ here also requires a unitary transformation in the state vector.

But Eugene, Johnny and Daniel each measured an *X*-magnitude of a laboratory plus atom simultaneously in $\Sigma$ when the quantum state of the laboratories plus atoms paralleled the GHZ state of three atoms but with $X_A$, $X_B$, $X_C$ in place of the *x*-spins of the atoms measured by Alice, Bob and Charlie respectively. Just as we saw in A.2 that the product of the outcome numbers *a, b, c* for measurements of *x*-spin must (with probability 1) be −1 in the GHZ state, so here the product of the outcome numbers *u, v, w* for measurements of $X_A$, $X_B$, $X_C$ respectively must (with probability 1) be −1, so $u.v.w. = -1$.[17] This is inconsistent with the previous three equations, as can be readily seen by multiplying them together to get $u.v.w. = +1$.

Note two things about the contradiction we have arrived at. Each of *a, b, c, u, v, w* numbers the single outcome of one quantum measurement actually recorded in this combined scenario: the scenario does not involve a sequence of repeated measurements of the same kind on similar systems, and nor does it involve any merely possible measurements. But repeated application of the Born rule in different frames assigns probability zero to every set of six numbers representing the outcomes of those six measurements. While a probability zero event may occur in an infinite event space (an infinitely thin dart may hit some point on a dartboard even though for each point there is zero probability of it's hitting at that point), the event space here is finite. The combined scenario presents us with a paradox insofar as the assumption of single outcomes is incompatible with the universal applicability of unitary quantum theory to isolated quantum systems.

## Acknowledgements

In addition to referees for this journal I wish to thank actual audiences at the University of Arizona and at STIAS; virtual audiences assembled by the Munich Centre for Mathematical Philosophy and also by the QISS project (Quantum Information Structure of Spacetime); and especially Nick Huggett, Tom Imbo, and Radwa Dawood for pressing me to clarify key points of section 4. I thank the Stellenbosch Institute for Advanced Study (STIAS) and the University of Arizona for their support.

Philosophy Department,
University of Arizona,
Tucson, Arizona, USA
rhealey@arizona.edu

---

17 Leegwater ([2018]) has $a.b.c. = -1$, which I consider a small slip or notational error.